\documentclass[letter,longauth,traditabstract]{aa} 
\usepackage{graphicx}
\usepackage{txfonts}
\usepackage{natbib}
\usepackage{epstopdf}
\def\revised{}

\usepackage{longtable}
\usepackage{rotating}
\usepackage{pdflscape}
\usepackage{multirow}

\begin{document}
    \title{Characterizing interstellar filaments with \emph{Herschel}\thanks{Herschel is an ESA space observatory 
    with science instruments provided by 
           European-led Principal Investigator consortia and with important participation from NASA.} in IC5146}
           

   \subtitle{}

   \author{
     D. Arzoumanian\inst{1}
      \and
            Ph. Andr\'e\inst{1}
        \and
            P. Didelon\inst{1}
         \and
          V. K\"onyves\inst{1}        
          \and
           N. Schneider\inst{1}
          \and
            A. Men'shchikov\inst{1}
             \and
             T. Sousbie\inst{2}
             \and
             A. Zavagno\inst{3}
          \and
          S. Bontemps\inst{4}
          \and
           J. Di Francesco\inst{5}
             \and
            M. Griffin\inst{6}           
            \and
            M. Hennemann\inst{1}      
            \and
            T. Hill\inst{1}
            \and
           J. Kirk\inst{6}
          \and
            P. Martin\inst{7}
            \and
            V. Minier\inst{1}
            \and
            S. Molinari\inst{8}                        
            \and
            F. Motte\inst{1}
             \and
            N. Peretto\inst{1}
            \and
            S. Pezzuto\inst{8}
            \and
            L. Spinoglio\inst{8}
             \and
            D. Ward-Thompson\inst{6}
             \and
          G. White\inst{9,11}
            \and
            C.D. Wilson\inst{10}             
            }

   \institute{Laboratoire AIM, CEA/DSM--CNRS--Universit\'e Paris Diderot, IRFU/Service d'Astrophysique, C.E.A. Saclay,
              Orme des Merisiers, 91191 Gif-sur-Yvette, France
              \email{doris.arzoumanian@cea.fr, pandre@cea.fr}
                \and
             Institute d'Astrophysique de Paris, UMR 7095 CNRS, Universit\'e Pierre et Marie Curie, 
             98 bis Boulevard Arago, F-75014 Paris, France
             \and 
             Laboratoire d'Astrophysique de Marseille, CNRS/INSU--Universit\'e de Provence, 13388 
             Marseille cedex 13, France
             \and
             Universit\'e de Bordeaux, OASU, Bordeaux, France
             \and          
             National Research Council of Canada, Herzberg Institute of Astrophysics,
             University of Victoria, Department of Physics and Astronomy, Victoria, Canada
              \and
             School of Physics \& Astronomy, Cardiff University, Cardiff, UK
              \and
             Canadian Institute for Theoretical Astrophysics, University of Toronto, Toronto, ON M5S 3H8, Canada
             \and
             IFSI - INAF, via Fosso del Cavaliere 100, I-00133 Roma, Italy
             \and
             The Rutherford Appleton Laboratory, Chilton, Didcot OX11 0NL, UK            
             \and
             Department of Physics and Astronomy, McMaster University, Hamilton, ON L8S 4M1, Canada 
             \and
             Department of Physics \& Astronomy, The Open University, Milton Keynes, UK 
                }
   \date{}

   \abstract{
{We provide a first look at the results of the $Herschel$ Gould Belt survey toward the IC5146 molecular cloud and present 
a preliminary analysis of the filamentary structure in this region.
The column density map, derived from our 70-500~$\mu$m $Herschel$ data, reveals a complex network of filaments, and confirms that 
these filaments are the main birth sites of prestellar cores.  
We analyze the column density profiles of 27 filaments and 
show that the underlying radial density profiles fall off as $r^{-1.5}$ to $r^{-2.5}$ at large radii. 
Our main result is that the filaments seem to be characterized by a narrow distribution of  widths having a median value of 
$0.10 \pm 0.03$~pc, which is in stark contrast to a much broader distribution of central Jeans lengths. This characteristic width of $\sim 0.1$~pc 
corresponds to within a factor of $\sim 2$ to the sonic scale below which interstellar turbulence becomes subsonic in diffuse gas, 
supporting the argument that the filaments may form as a result of the dissipation of large-scale turbulence.
 }

   \keywords{ISM: individual objects (IC5146) -- Stars: formation  -- ISM: clouds    -- ISM: Filaments
                -- ISM: structure -- submillimeter 
             }}
\date{Accepted: 17/02/2011}

   \maketitle
%

\section{Introduction}

Understanding how stars form out of the diffuse interstellar medium (ISM) 
on both global and local scales is a fundamental open problem in contemporary astrophysics 
\citep[see][for a recent review]{McKee2007}.
%
Much observational progress is being made on this front thanks to the $Herschel$ Space Observatory 
\citep{Pilbratt2010}.
%
%
In particular, the first results from the Gould Belt and Hi-GAL imaging surveys with $Herschel$ have revealed 
a profusion of parsec-scale 
filaments in Galactic molecular clouds 
and suggested an intimate connection between the filamentary structure 
of the ISM and the formation process of dense cloud cores \citep{Andre2010, Molinari2010}. 
Remarkably, filaments are omnipresent even in unbound, non-star-forming complexes such as the Polaris translucent cloud 
\citep{Men'shchikov2010, Miville2010, Ward-Thompson2010}. 
Furthermore, in active star-forming regions such as the Aquila Rift cloud, most of the prestellar cores identified with $Herschel$ 
are located within gravitationally unstable filaments for which the mass per unit length exceeds the 
critical value 
\citep[][]{Ostriker1964}, $M_{\rm line, crit} = 2\, c_s^2/G \sim 15\, M_\odot$/pc,  
where $c_{\rm s} \sim 0.2$~km/s is the isothermal sound speed for $T \sim 10$~K. 
The  early findings from $Herschel$ led \citet{Andre2010} to favour a scenario according to which core formation 
occurs in two main steps. First, large-scale magneto-hydrodynamic (MHD) turbulence generates a whole network of 
filaments in the ISM \citep[cf.][]{Padoan2001,Balsara2001}; second, the densest filaments fragment into 
prestellar cores by  
gravitational instability \citep[cf.][]{Inutsuka1997}.

To refine this observationally-driven scenario of core formation and gain insight into the origin of the 
filamentary structure, an important step is to characterize the detailed physical properties of the filaments seen 
in the $Herschel$ images. 
Here, we present new results from the Gould Belt survey obtained toward the IC5146 molecular cloud
and analyze the radial density profiles of the numerous filaments identified in this cloud.
Based on a comparison with a similar analysis for the Aquila and Polaris regions, we show that the filaments of IC5146, Aquila, and 
Polaris are all characterized by a typical width of $\sim 0.1$~pc and discuss possible physical implications of this finding. 

IC5146 is a star-forming cloud in Cygnus located at a distance of $\sim 460$~pc \citep[][but see Appendix A]{Lada1999}, which consists of the Cocoon Nebula, an H\,{\sc{ii}} region  
illuminated by the B0 V star  $BD+46^{\circ} 3474$, and two parallel filamentary streamers extending to the west \citep{Lada1994}.
In addition to these two streamers, the $Herschel$ images reveal a whole network of filaments, which are the focus of the present letter.



   \begin{figure*}
   \centering
     \resizebox{9.cm}{!}{
   \includegraphics[angle=0]{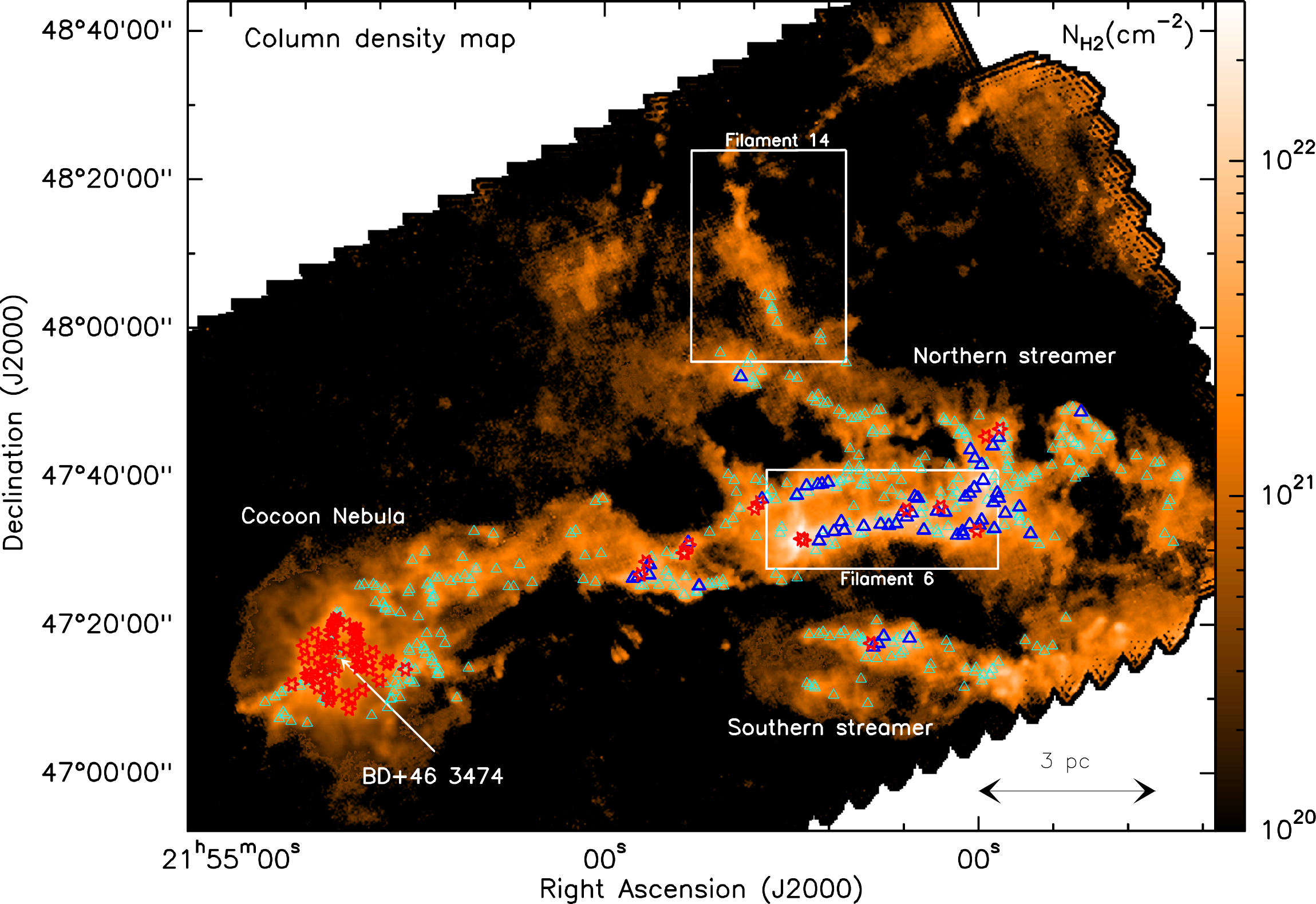}}
      \hspace{0.4mm}
  \resizebox{9.cm}{!}{
  \includegraphics[angle=0]{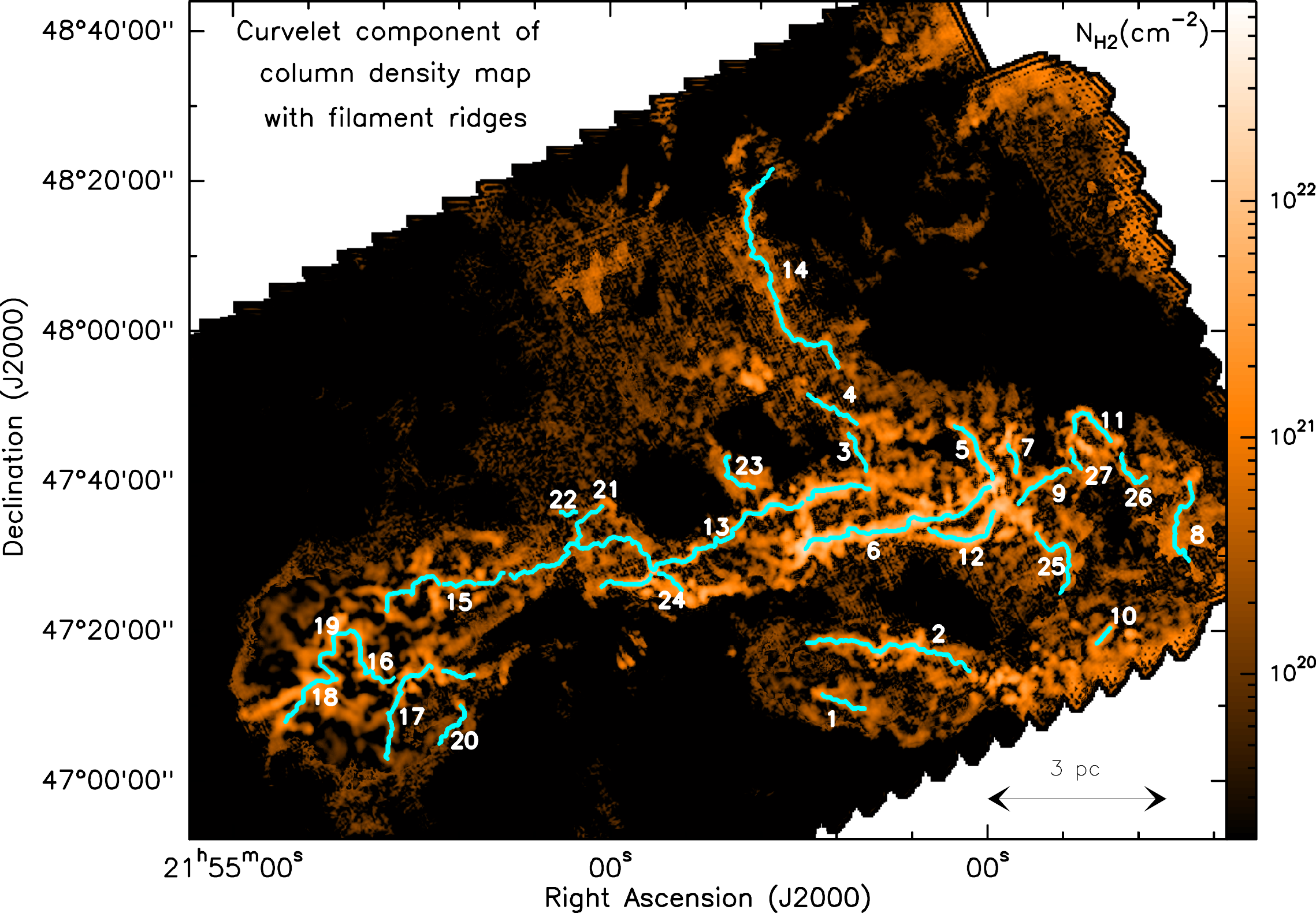}}
  \caption{{\bf (a)} Column density map derived from our SPIRE/PACS observations of IC5146. The resolution corresponds to 
the $36.9\arcsec $~HPBW resolution of SPIRE at 500~$\mu$m. 
The positions of the YSOs and starless cores detected with $Herschel$ \citep[using {\it getsources}, a source-finding algorithm described in][]{Men'shchikov2010} 
are plotted as red stars and blue triangles respectively; bound starless cores are in dark blue 
\citep[cf. Fig.~\ref{250_curv_crit}b and][for classification details]{Konyves2010}.
The locations of the two filaments (6 and 14) whose radial  profiles are shown in Fig.~\ref{fil_prof} and online Fig.~\ref{fil_prof_14} 
are marked by the white rectangles. {\bf (b)} {\it Curvelet} component of the column density map,
with the network of 27 identified filaments shown in blue.
}
              \label{coldens}
    \end{figure*}


\section{Observations and data reduction}

Our observations of IC5146  were made on 29 May 2010 in the parallel scan-map mode of $Herschel$. 
An area of $\sim$1.6~deg$^2$ was covered by both PACS  \citep{Poglitsch2010}  at 70~$\mu$m, 160~$\mu$m and  
SPIRE  \citep{Griffin2010}  at 250~$\mu$m, 350~$\mu$m, 500~$\mu$m,  with a scanning speed of $60\arcsec \, {\rmÊs}^{-1}$. 
%
The PACS data were reduced with HIPE version 3.0.1528. Standard
steps of the default pipeline were applied, starting from the raw data, 
taking special care of the deglitching and high-pass filtering steps. 
The final  maps were created using the photProject task. 
%
The SPIRE observations 
were reduced using
HIPE version 3.0.1484.
The pipeline scripts were modified to include data taken
during the turnaround of the telescope. A median
baseline was applied and  the 'naive' map-making method was used. 
%
     %
     


Thanks to their unprecedented spatial dynamic range in the submillimeter regime, the $Herschel$ images 
(see, e.g., online Fig.~\ref{color_image} and Fig.~\ref{250_curv_crit}) provide 
a wealth of detailed quantitative information 
on the large-scale, filamentary structure of the cloud.  
It is this filamentary structure that we discuss in the following. 
 %
 
%
\section{Analysis of the filamentary structure}
%


\subsection{Identification of a network of filaments}

Based on the {\it Herschel} images at five wavelengths, {\revised a dust temperature ($T_{\rm d}$) map 
(see online Fig.~\ref{250_curv_crit}a) }
and a column density ($ N_{\rm H_{2}}$)
map (Fig.~\ref{coldens}a) were constructed for IC5146, assuming the dust opacity law 
of \citet[see Appendix A]{Hildebrand1983} 
and following the same procedure as \citet{Konyves2010} for Aquila.
%
We decomposed the column density map on curvelets and wavelets \citep[e.g.][]{Stark2003}. 
The curvelet component image (Fig.~\ref{coldens}b) 
provides a high-contrast view 
of the filaments 
(after subtraction of dense cores,  which are contained in the wavelet component).
%
We then applied the DisPerSE algorithm \citep{Sousbie2010} to the curvelet image 
to take a census of the filaments and trace their ridges. 
DisPerSE is a general method 
to identify structures such as filaments and voids 
in astrophysical data sets (e.g. gridded maps). 
The method, based on principles of computational topology, traces filaments by connecting saddle points 
to maxima with integral lines.  
{\revised  From the filaments identified with DisPerSE above a 'persistence' threshold 
of $5\times 10^{20}{\rm cm}^{-2}$ in column density 
($\sim 5\sigma $ in the curvelet map), }
we built a mask on the same grid as the input map, with values of 1 along the filament
ridges and 0 elsewhere {\revised \citep[cf.][for a definition of 'persistence']{Sousbie2010}. }
Based on the skeleton mask obtained for IC5146, we identified and numbered a total of 
27 filaments (shown in  blue in Fig.~\ref{coldens}b).

%


  \begin{figure*}
   \centering
     \resizebox{9.cm}{!}{
\includegraphics[angle=0]{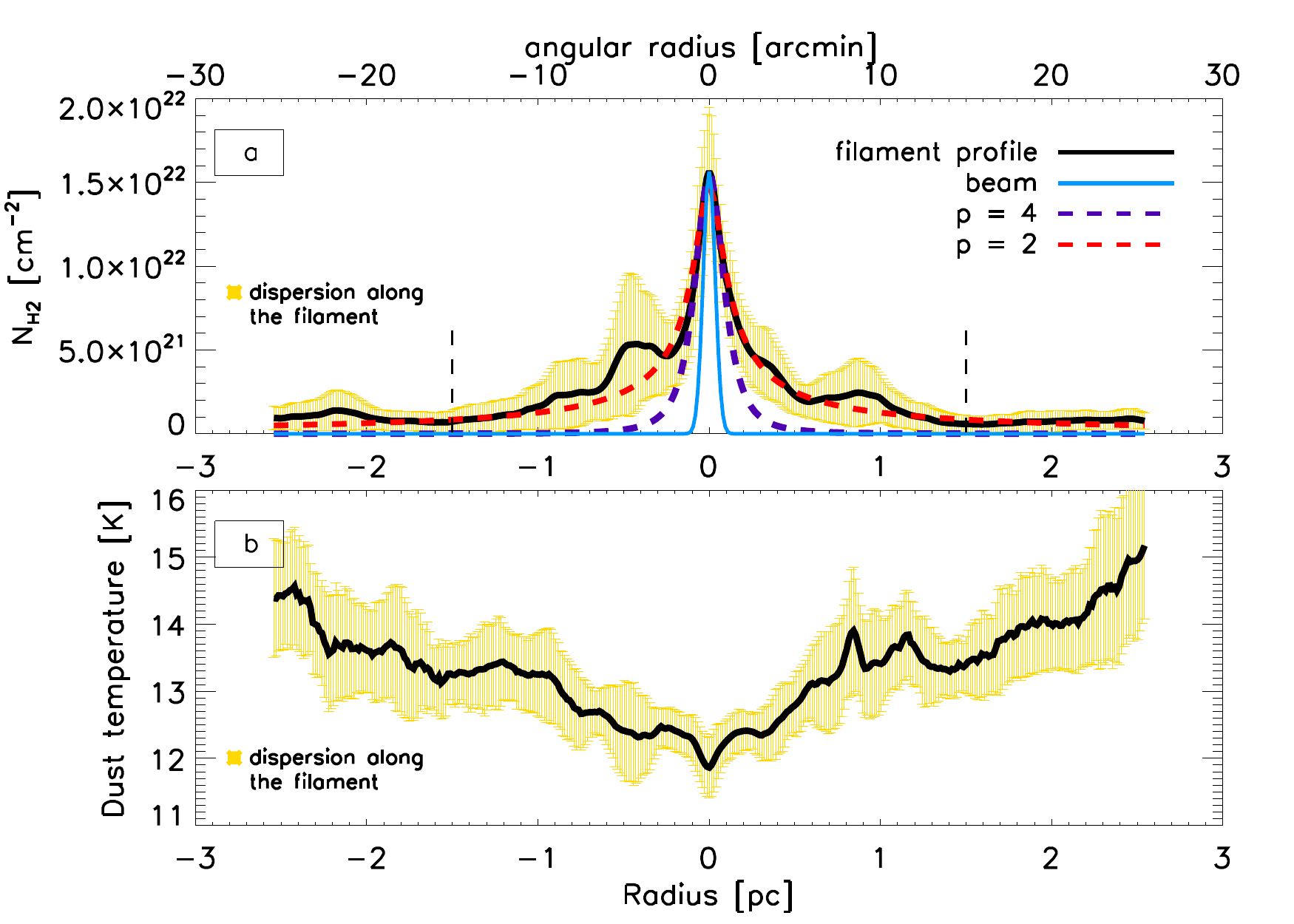}}
      \hspace{2mm}
  \resizebox{8.8cm}{!}{
 \includegraphics[angle=0]{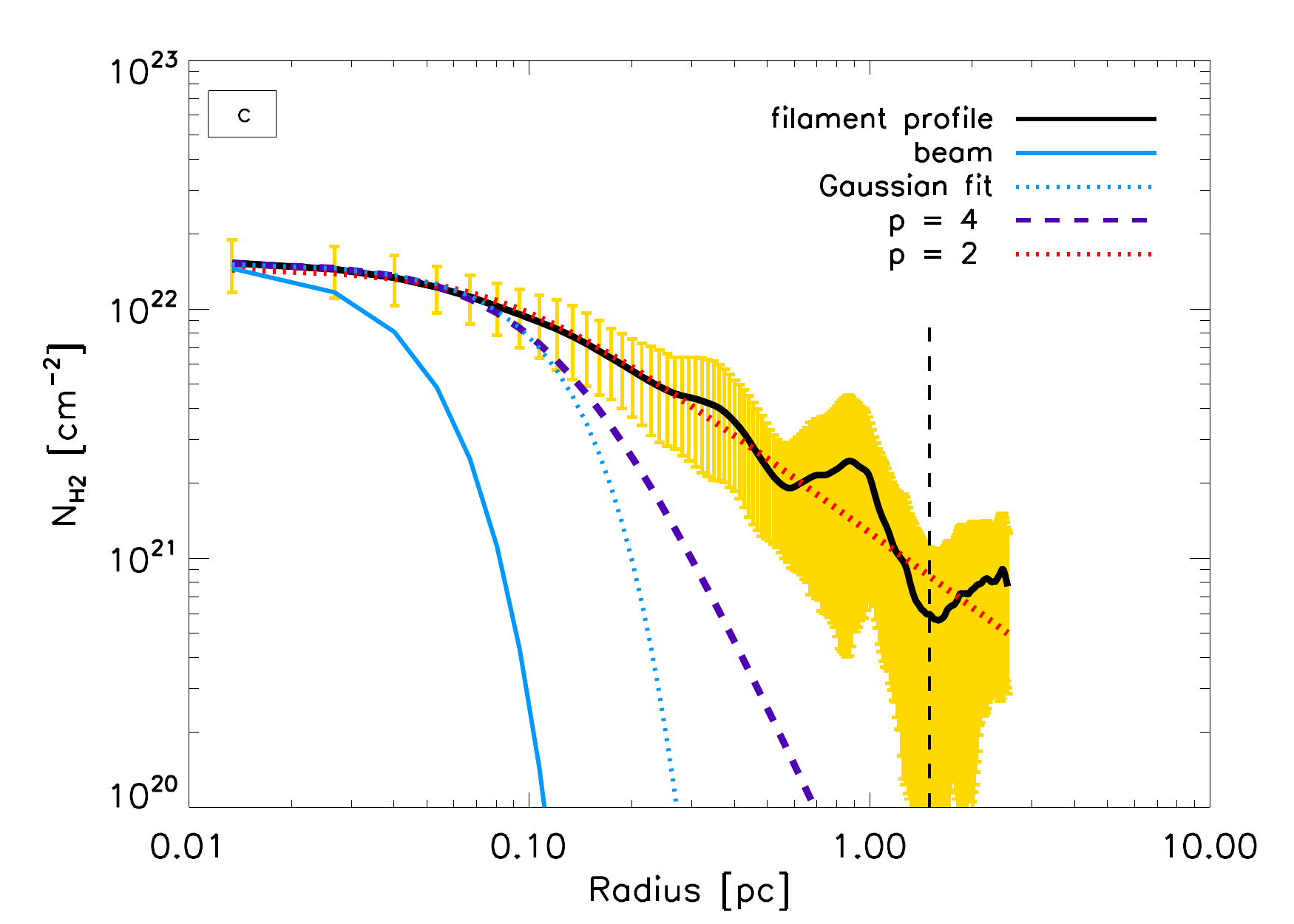}}
   \caption{{\bf(a)} Mean radial column density profile perpendicular to the supercritical filament  labeled 6
in Fig.~\ref{coldens}b (black curve). The area in yellow shows the dispersion 
of the radial profile along the filament. 
The inner curve in light blue corresponds to the effective 36.9\arcsec ~HPBW resolution
(0.08~pc at 460~pc) of  the column density map of Fig.~\ref{coldens}a used to construct the profile.
The dotted red curve shows the best-fit model of the form expressed by Eq.~(1), with $p = 2$.
For comparison, the dashed blue curve shows a model with $p = 4$, corresponding to a hydrostatic isothermal
equilibrium filament \citep{Ostriker1964}. The two dashed vertical segments at $\sim \pm 1.5$~pc mark the bounds
within which the profile was integrated to estimate the mass per unit length $M_{\rm line}$.
{\bf(b)} Mean radial dust temperature profile measured perpendicular to filament 6.  
{\bf(c)} Same as {\bf(a)}  displayed in log-log format to better visualize the flatter inner part of the profile and the power-law behavior at large radii. }
              \label{fil_prof}
    \end{figure*}
    

\subsection{Radial density profiles of the filaments}

To construct the mean radial density profile of each filament, the tangential direction to the filament's ridge was first computed at each pixel position 
along the filament. Using the {\it original} column density map (cf. Fig.~\ref{coldens}a), we then measured a radial column density profile 
perpendicular to the ridge at each position. Finally, we derived the mean radial profile by averaging all of the profiles along the filament (see Fig.~\ref{fil_prof}).
%
%
In order to characterize each observed (column) density profile, we adopted an idealized model of a cylindrical filament  
with radial density and column density profiles (as a function of cylindrical radius $r$) of the form: 
\vspace{-0.2cm}
$$ \rho_{p}(r) = \frac{\rho_{c}}{\left[1+\left({r/R_{\rm flat}}\right)^{2}\right]^{p/2}}\  \longrightarrow 
 \Sigma_{p}(r) = A_{p}\,  \frac{\rho_{\rm c}R_{\rm flat}}{\left[1+\left({r/R_{\rm flat}}\right)^{2}\right]^{\frac{p-1}{2}}}, \ \ (1)$$
where $\Sigma = \mu m_{\rm H} N_{\rm H_{2}}$, $\mu=2.33$ is the mean molecular mass, 
$ A_{p}$~=~$ \frac{1}{\cos\,i} \, \int_{-\infty}^{\infty}\frac{du}{(1+u^{2})^{{p}/2}} $
is a finite constant factor  for $p>1$, $\rho_{\rm c}$ is the density at the center of the filament, and $R_{\rm flat}$ is 
the characteristic radius of the flat inner portion of the density profile\footnote{Model profiles of the form described by Eq.~(1) are sometimes 
called Plummer-like \citep[cf.][and references therein]{Nutter2008}.
When $R_{\rm flat}$ is much smaller than the spatial resolution of the observations, such profiles are effectively equivalent to power laws.}.
For simplicity, the inclination angle $i$ of the model filament to the plane of the sky is assumed here to be $i = 0$ (but see online 
Appendix~\ref{app_cosi} for the effect of $i \neq 0$).
At large radii ($r >> R_{\rm flat}$), the model density profile approaches a power law: $\rho_{{p}}(r) \sim \rho_{c} \left(r/R_{\rm flat}\right)^{-p}$. 
An important special case is the \citet{Ostriker1964} model of an isothermal filament in hydrostatic equilibrium, for which 
$p = 4$, $ A_{p}= \pi/2$, and  $R_{\rm flat}$ corresponds to the thermal Jeans length at the center of the filament, i.e.,  
$R_{\rm flat} \equiv  \lambda_{\rm J}\left({r=0}\right) =  c_{\rm s}^{2}/\left(G \Sigma_{r=0}\right) $.

When fitting such a model profile to the observations, 
 $\rho_{\rm c}$, $R_{\rm flat}$, and $p$, along with the peak position of the filament, were treated as free parameters.
The results (see, e.g., Fig.~\ref{fil_prof} and online Table~1) show that the observed profiles are generally well fitted with $1.5 < p < 2.5$. 
None of the observed filaments has the steep $p = 4$ density profile of the \citet{Ostriker1964} model.
A similar conclusion was already reported by \citet{Lada1999} in the case of the main streamer (i.e., Filament 6 here).

The above radial profile analysis can also be used to derive accurate masses per unit length, $M_{\rm line}$, for the filaments 
by integrating column density over radius: $M_{\rm line}= \int \Sigma_{\rm obs}(r) {\rm d}r$. 
The main source of uncertainty lies in the difficulty of defining the edges of the filaments, especially in crowded regions. 
The method is nevertheless quite robust when the bounds of integration are reasonably well defined (cf. Fig.~\ref{fil_prof}). 
The values of $M_{\rm line}$ derived with this method (see online Table~1) 
are typically $20\%$ higher than simpler estimates 
assuming that the filaments have Gaussian column density profiles \citep[cf. Appendix of][]{Andre2010}.
Most bound prestellar cores appear to be located within supercritical, gravitationally unstable  
filaments with $M_{\rm line} > M_{\rm line,crit}$ 
(e.g. Filament 6 
-- see online Fig.~\ref{250_curv_crit}b), a similar result to that already obtained by \citet{Andre2010} in Aquila. 
Note that both methods of estimating $M_{\rm line}$ (profile integration or Gaussian approximation) yield essentially the same 
conclusion on 
the gravitational instability of the filaments. 


\section{A characteristic width for interstellar filaments ?}

To construct a reliable distribution of filament widths (cf. online Fig.~\ref{histo}a), we applied   Gaussian fits 
to the radial column density profiles, as these tend to be more robust than the more sophisticated fits discussed in Sect.~3.2. 
A Gaussian fit to a model profile of the form of Eq.~(1) with $p = 2$ indicates that the derived FWHM width 
is equivalent to $\sim 1.5 \times (2 \, R_{\rm flat})$. 
%
It can be seen in Fig.~\ref{histo}a that the filaments of IC5146 have a narrow distribution of 
deconvolved FWHM widths centered around a typical (median) value of $0.10 \pm 0.03$~pc (see also Fig.~\ref{width_coldens}). 
Note that the same filaments span more than an order of magnitude in central column density (cf. online Table~1), 
implying a distribution of central Jeans lengths [$ \lambda_{\rm J}\left({r=0}\right) =  c_{\rm s}^{2}/\left({G \Sigma_{r=0}}\right) $]  
from $\sim 0.02$~pc to $> 0.3$~pc, which is much broader than the observed distribution of  widths 
(see Fig.~\ref{width_coldens} and blue dashed line in Fig.~\ref{histo}a).


In order to check that the measured filament widths were not strongly affected by the finite resolution 
of  the column density map, {\revised we performed the same radial profile analysis using the SPIRE 250~$\mu$m map, 
which has a factor $\sim 2$ better resolution (18.1\arcsec~HPBW). 
The resulting distribution of filament widths is shown as a dotted histogram 
in online Fig.~\ref{histo}a and is statistically very similar to the
original distribution at 36.9\arcsec ~resolution (at the 50\% confidence level according to a Kolmogorov-Smirnov test).}



We also performed a similar analysis of the filamentary structure in two other regions 
located 
at different distances, 
the Aquila and Polaris fields, also observed by us with $Herschel$. 
The Aquila region is a very active star-forming complex at $d \sim 260$~pc 
\citep[e.g. ][and references therein]{Bontemps2010}, while the Polaris field is a high-latitude {\it translucent} cloud
with little to no star formation at $d \sim 150$~pc \citep[e.g.][and references therein]{Ward-Thompson2010}.
Following the same procedure as in Sect.~3.1, we identified 32 filaments in Aquila and 31 filaments in Polaris 
spanning three orders of magnitude in central column density from $\sim 10^{20}$~c$\rm{m}^{-2}$ for the most 
tenuous filaments of Polaris to $\sim 10^{23}$~c$\rm{m}^{-2}$ for the densest filaments of Aquila 
\citep[see also Fig.~1 of ][]{Andre2010}. The distribution of  deconvolved FWHM widths 
for the combined sample of 90 filaments 
is shown in online Fig.~\ref{histo}b.
This combined distribution is 
sharply peaked at $0.10\pm 0.03$~pc, which strengthens the trend noted above for IC5146. 
While further tests would be required to fully investigate potential biases 
(especially given uncertainties in cloud distances -- see Appendix~A) and reach a definitive conclusion, 
our present findings suggest that most interstellar filaments may share a similar characteristic width of $\sim 0.1$~pc.

\section{Discussion} 

\begin{figure}
    \begin{minipage}{1\linewidth}
     \resizebox{\hsize}{!}{\includegraphics[angle=0]{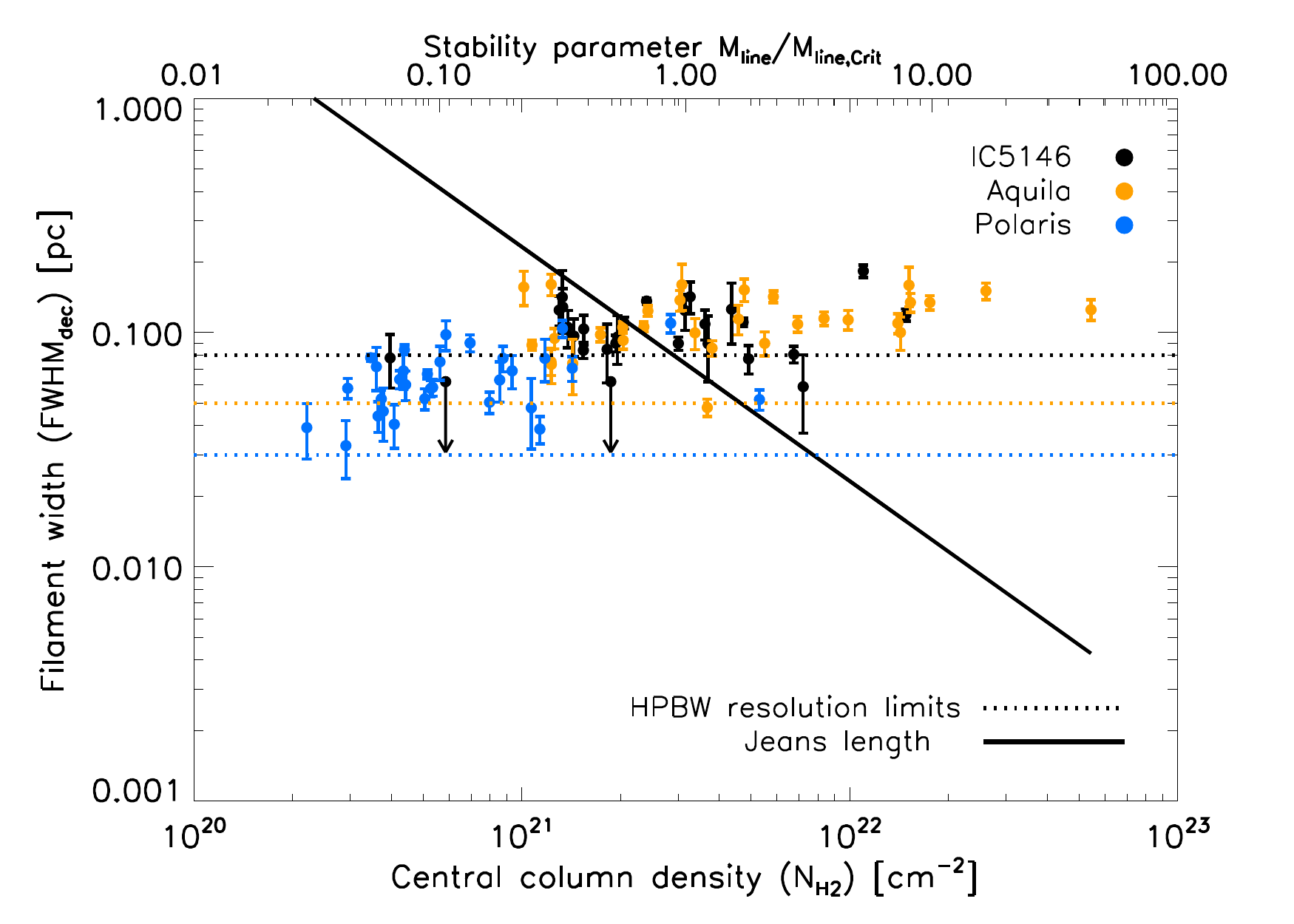}}
       \end{minipage}   
   \caption{Mean deconvolved width (FWHM) versus mean central column density for the 27 filaments of IC5146 shown in Fig~\ref{coldens}b, 
along with 32 filaments in Aquila and 31 filaments in Polaris, all analyzed in the same way (see Sect.~3).  
The spatial resolutions of the column density maps used in the analysis for the three regions are marked by the horizontal dotted lines. 
The filament width has a typical value of $\sim 0.1$~pc, regardless of central column density. 
The solid line running from top left to bottom right shows the central (thermal) Jeans length as a function of central column density 
[$\lambda_{J} =  c_{s}^{2}/\left({G \Sigma_{0}}\right) $] for a gas temperature of 10~K.}
   \label{width_coldens}%
    \end{figure}

The results presented in Sect.~3 and Sect.~4  can be used to discuss the formation and evolution 
of the observed filamentary structure.  
Three broad classes of models have been proposed in the literature to account for filaments in molecular clouds, 
depending on whether global gravity \citep[e.g.][]{Heitsch2008}, 
magnetic fields \citep[e.g.][]{Nakamura2008}, or large-scale turbulence \citep[e.g.][]{Padoan2001,MacLow2004} 
play the dominant role.
Large-scale gravity can hardly be invoked to form 
{\revised  filaments in gravitationally unbound complexes such as   the Polaris flare cloud  \citep[cf.][]{Heithausen1990}}.
While magnetic fields may play an important role in channeling mass accumulation onto the densest filaments 
(e.g. Goldsmith et al. 2008, Nakamura \& Li 2008), 
our $Herschel$ findings appear to be consistent with the turbulent picture. 
In the scenario proposed by \citet{Padoan2001}, the filaments
correspond to dense, post-shock stagnation gas associated with regions of converging supersonic flows. 
One merit of this scenario is that it provides an explanation for the typical $\sim 0.1$~pc width of the filaments as measured 
with $Herschel$ (Sect.~4 and Fig.~\ref{width_coldens}). In a plane-parallel shock, the thickness $\lambda $
of the postshock gas layer/filament is related to the thickness $L$ of the preshock gas by $\lambda \approx L/\mathcal{M}(L)^2 $, 
where $\mathcal{M}(L)$ is the Mach number and $\mathcal{M}(L)^2 $ is the compression ratio of the shock 
for a roughly isothermal radiative hydrodynamic shock. 
Thus, the thickness $\lambda $
is independent of the scale $L$ given Larson's linewidth--size 
relation 
[$\mathcal{M}(L) \propto \sigma_{\rm v}(L) \propto L^{0.5} $]. 
In this picture, the postshock thickness of the filaments effectively corresponds to the sonic scale $\lambda_{\rm s}$ at  
which the 3D turbulent velocity dispersion equals the sound speed (i.e., $\mathcal{M}(\lambda_{\rm s}) = 1$), leading 
to $\lambda \approx \lambda_{\rm s} \sim $~0.05--0.15~pc according to recent determinations of the linewidth--size 
relationship in molecular clouds \citep[e.g.][]{Heyer2009, Federrath2010}.

If large-scale turbulence provides a plausible mechanism for {\it forming} the filaments, the fact that prestellar cores 
tend to form in gravitationally unstable filaments \citep[][]{Andre2010} suggests that gravity is a major driver 
in the subsequent {\it evolution} of the filaments. The power-law shape of the outer density profiles, especially in the 
case of supercritical filaments (Sect.~3.2 and Fig.~\ref{fil_prof}), is also suggestive of the role of gravity. 
Although the observed profiles are shallower than the profile of a self-gravitating isothermal 
equilibrium filament  \citep{Ostriker1964}, they are consistent with some models of magnetized filaments in gravitational 
virial equilibrium \citep{Fiege2000}.  
Another attractive explanation of the observed $\rho \sim  r^{-2}$  profiles is that the filaments are not strictly isothermal 
and that some of them are collapsing. \citet{Nakamura1999} have shown that the cylindrical 
version of the Larson-Penston similarity solution for the collapse of a filament has an outer density profile which 
approaches 
$\rho \propto r^{-2}$ when the equation of state is not isothermal but polytropic 
($P \propto \rho^\gamma $) with $\gamma \la 1$. 
The dust temperature profiles derived for the IC5146 filaments generally show a slight, but significant 
temperature drop toward the center of the filaments (cf. Fig.~\ref{fil_prof}b), suggesting that a polytropic 
equation of state with $\gamma \la 1$ may indeed be more appropriate than a simple isothermal assumption.

The lack of anti-correlation between filament width and central column density (see Fig.~\ref{width_coldens}) provides another 
strong constraint on filament evolution. 
If the filaments are initially formed with a characteristic thickness $\sim 0.1$~pc as a result 
of turbulent compression (see above), then the left-hand (subcritical) side of Fig.~\ref{width_coldens} 
can be readily understood. 
The right-hand 
(supercritical) 
side 
of Fig.~\ref{width_coldens} 
is more surprising since one would naively 
expect an anti-correlation between width and central column density (e.g. $W \propto 1/\Sigma_0 $) 
for self-gravitating filaments contracting at fixed mass per unit length. 
However, the velocity dispersion $\sigma_{ \rm v}$ and
the virial mass per unit length, $M_{\rm line, vir} = 2\, \sigma_{ \rm v}^2/G$ 
(replacing $M_{\rm line, crit}$ in the presence of nonthermal motions -- see Fiege \& Pudritz 2000),  
may increase during filament contraction as gravitational energy is converted into kinetic energy. 
If approximate virial balance is maintained, the width $ \sim {\sigma_{\rm v}^{2}}/\left({G \Sigma_{0}}\right) $ 
may thus remain nearly constant as in Fig.~\ref{width_coldens}. 
A trend such as $ \sigma_{\rm v} \propto \Sigma ^{0.5} $ has indeed been 
found by \citet{Heyer2009} in their study of velocity dispersions in Galactic molecular clouds. 
To account for the right-hand side of Fig.~\ref{width_coldens}, we thus hypothesize that gravitationally unstable filaments 
may accrete additional mass from their surroundings and {\it increase in mass per unit length} while contracting. 
Interestingly, mass accretion onto the densest filaments of Taurus and Cygnus X may have been observed in the form 
of CO striations aligned with the local magnetic field \citep{Goldsmith2008,Nakamura2008, Schneider2010}.
{\revised Molecular line observations \citep[cf.][]{Kramer1999} }
of a broad sample of $Herschel$ filaments in several regions would be extremely valuable 
to confirm the validity of the scenario proposed here.

%
%

\onlfig{5}{    
   \begin{figure*}
   \centering
\hspace{-0.5cm}  \resizebox{9.5cm}{!}{ 
   \includegraphics[angle=0]{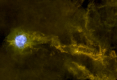}}
      \hspace{0.75cm}  \resizebox{8.25cm}{!}{   \includegraphics[angle=0]{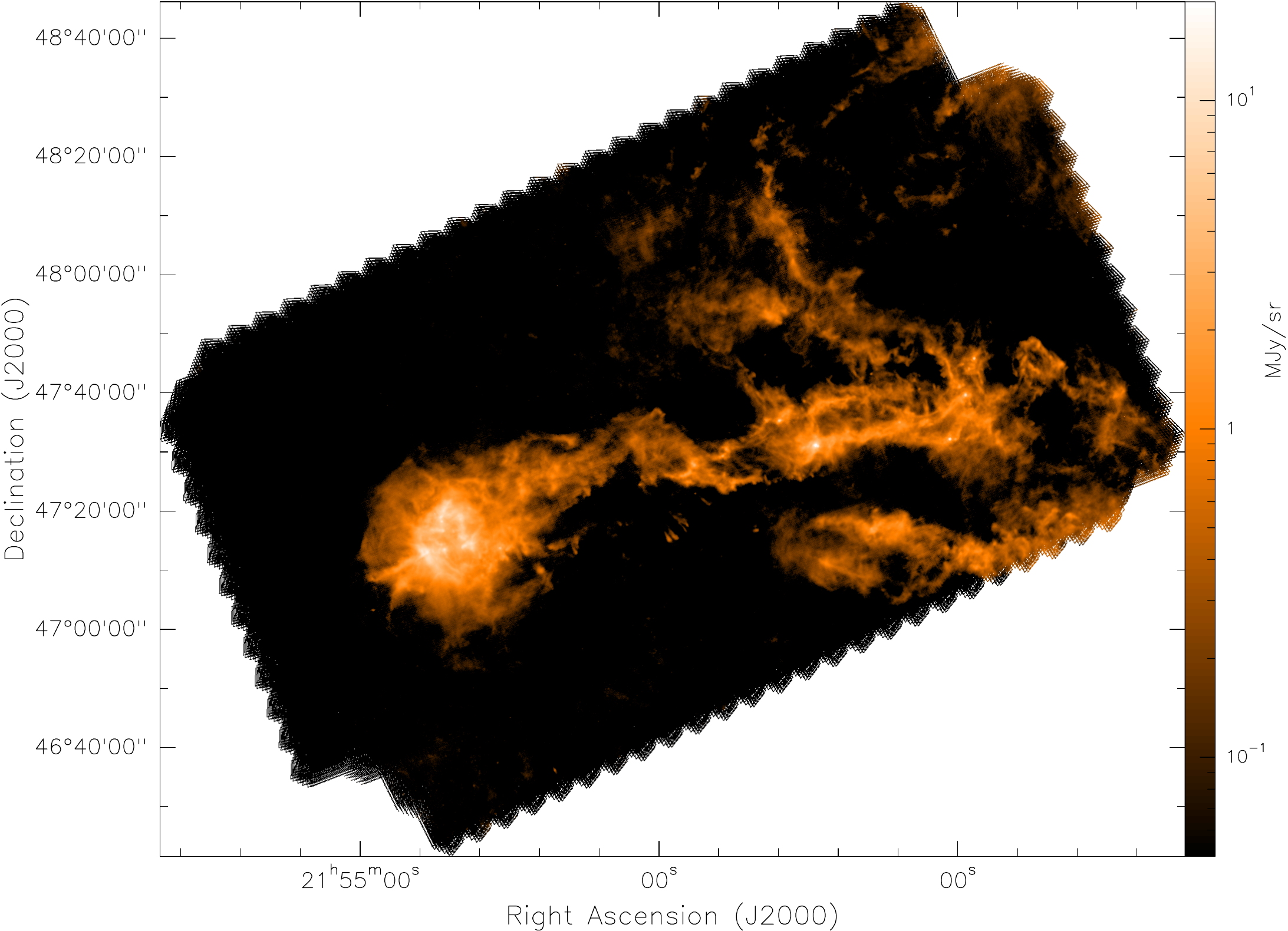}}
  \caption{{\bf(a)} Composite 3-color image of IC5146 ($\sim 1.6$~deg$^2$ field) produced from our PACS/SPIRE parallel-mode data at 70, 250, and $500 \mu$m. 
The color coding is such that red = SPIRE 500 $\mu $m,  green = SPIRE 250 $\mu$m, blue = PACS 70~$\mu$m. 
{\bf(b)} SPIRE 250~$\mu$m image of IC5146.
}            \label{color_image}%
    \end{figure*}
    } 
    
\onlfig{6}{
   \begin{figure*}
   \centering
   \vspace{0.75cm}
     \resizebox{8.75cm}{!}{
   \includegraphics[angle=0]{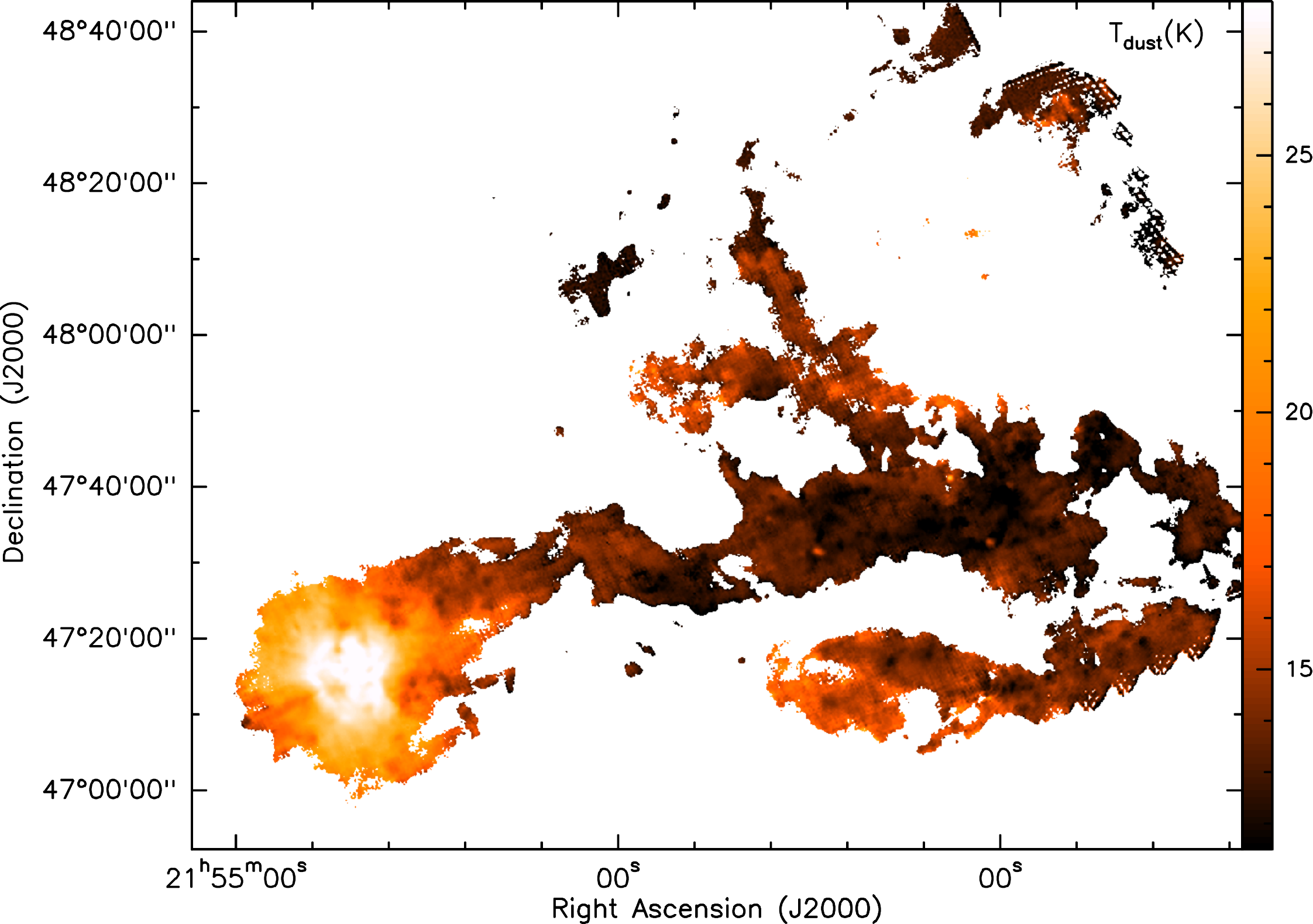}}
      \hspace{4mm}
  \resizebox{9.cm}{!}{
  \includegraphics[angle=0]{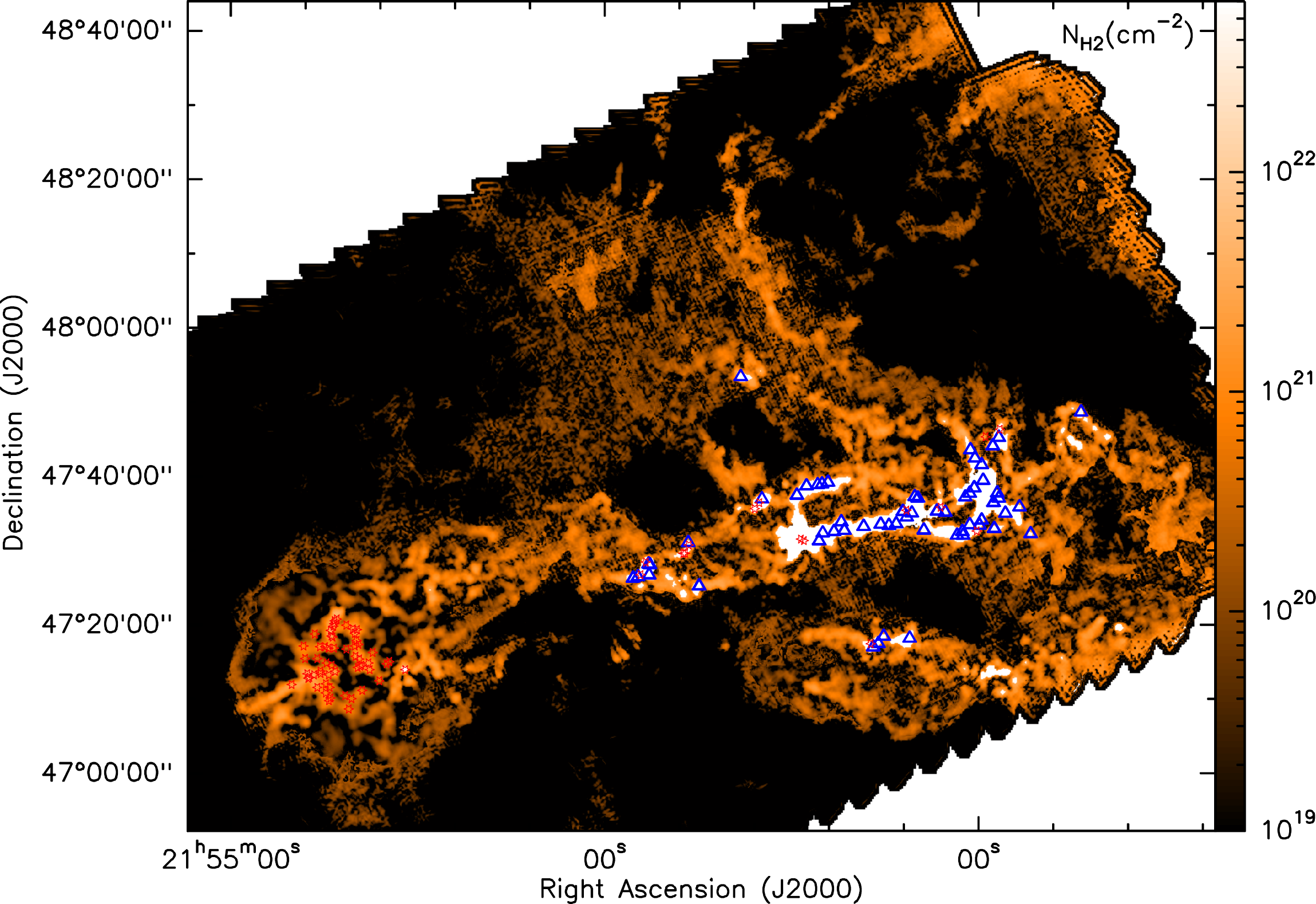}}
  \caption{{\bf(a)} {\revised Dust temperature map derived from our SPIRE/PACS observations of IC5146. The resolution corresponds 
to the $36.9\arcsec $~HPBW resolution of SPIRE at 500~$\mu$m. }
  {\bf(b)} Curvelet component of the column density map of IC5146 (cf. Fig.~\ref{coldens}), in which 
  the areas where the filaments have a mass per unit length larger than half the critical value and are thus likely gravitationally unstable have been highlighted in white.
  The positions of the 71 YSOs and 45 candidate bound prestellar cores identified  with {\it getsources}  
  \citep[][]{Men'shchikov2010} in the $Herschel$ images are overplotted as red stars and dark blue triangles, respectively.
  Candidate bound prestellar cores were selected among a larger population of starless cores (see blue triangles in Fig.~\ref{coldens}a), 
  on the basis of a comparison of the core masses with local values of the Jeans or Bonnor-Ebert mass
   \citep[see Sect.~4.1 of][ for details]{Konyves2010}. 
  Note the excellent correspondence between the spatial distribution of the prestellar cores and the supercritical filaments, in agreement with our earlier results  in
  Aquila and Polaris \citep[see Fig.~2 of][]{Andre2010}.}
              \label{250_curv_crit}
    \end{figure*}
    }
    
\onlfig{4}{   
   \begin{figure}
   \centering
  \resizebox{9.cm}{!}{
   \includegraphics[angle=0]{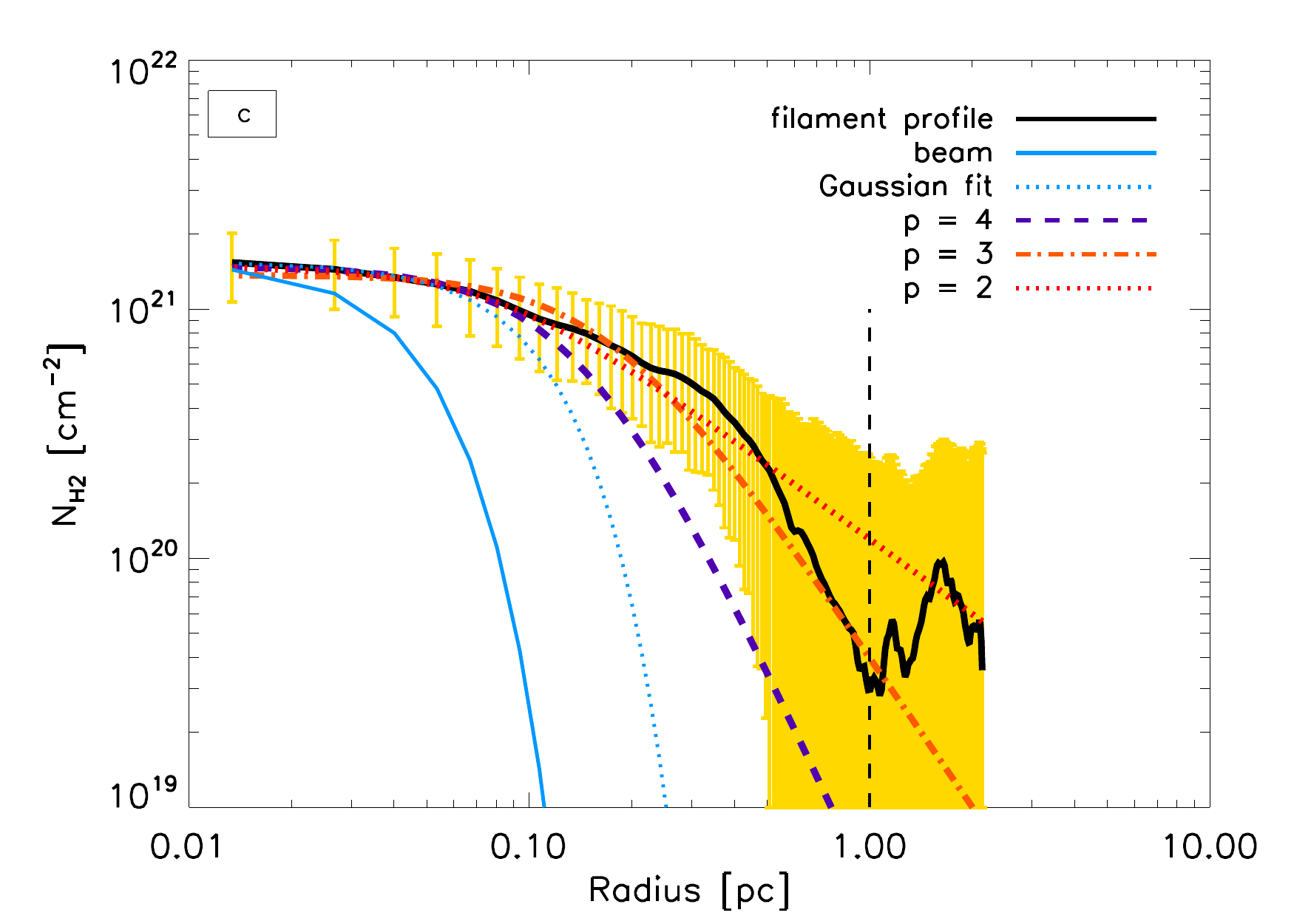}}
  \caption{Mean radial column density profile in log-log format of the subcritical filament labeled 14 in Fig.~\ref{coldens}b (western side).
Note that, given the relatively large dispersion 
of the radial profile along the filament (shown in yellow), the power-law behavior at large radii is less clear 
in this subcritical case than for the supercritical filament shown in Fig.~\ref{fil_prof}c.  }           
   \label{fil_prof_14}
    \end{figure}
    }

\begin{acknowledgements}
SPIRE has been developed by a consortium of institutes led by
Cardiff Univ. (UK) and including Univ. Lethbridge (Canada);
NAOC (China); CEA, LAM (France); IFSI, Univ. Padua (Italy);
IAC (Spain); Stockholm Observatory (Sweden); Imperial College
London, RAL, UCL-MSSL, UKATC, Univ. Sussex (UK); Caltech, JPL,
NHSC, Univ. Colorado (USA). This development has been supported
by national funding agencies: CSA (Canada); NAOC (China); CEA,
CNES, CNRS (France); ASI (Italy); MCINN (Spain); SNSB (Sweden);
STFC (UK); and NASA (USA).  
PACS has been developed by a consortium of institutes led by MPE
(Germany) and including UVIE (Austria); KUL, CSL, IMEC (Belgium); CEA,
OAMP (France); MPIA (Germany); IFSI, OAP/AOT, OAA/CAISMI, LENS, SISSA
(Italy); IAC (Spain). This development has been supported by the funding
agencies BMVIT (Austria), ESA-PRODEX (Belgium), CEA/CNES (France),
DLR (Germany), ASI (Italy), and CICT/MCT (Spain).

\end{acknowledgements}


\onltab{1}{
\begin{table*}  
\label{table_param}    
\centering
 \caption{Summary of derived parameters for the 27 filaments of IC5146}
\begin{tabular}{|c|cc|cc|cc|cc|c|c|c|  }   
\hline\hline   
Filament    & $N^0_{\rm H_{2}} $ &$\sigma_{N_{\rm H_{2}} }$$^{\bf(a)}$ &  $T^0_{\rm d}$ &$\sigma_{\rm T_{\rm d}}$$^{\bf(a)}$& $\langle\,{\rm FWHM}_{\rm dec}\rangle$   & $\sigma_{\langle{\rm FWHM}_{\rm dec}\rangle}$$^{\bf(a)}$ &$\lambda^0_{\rm J}$ &$\sigma_{\lambda_{\rm J}}$  $^{\bf(a)}$&$p$\,$^{\bf(b)}$&$R_{\rm flat}$$^{\bf(b)}$&$M_{\rm line}$$^{\bf(c)}$\\ 
 & \multicolumn{2}{|c|}{[$10^{21}$ c$\rm m^{-2}$] }  & \multicolumn{2}{|c|}{ [K] }& \multicolumn{2}{|c|}{[pc]}&   \multicolumn{2}{|c|}{[pc]}& &[pc]& $[M_{\odot}$/pc]\\
  (1) & (2)  & (3)&(4)&(5)&(6)&(7)&(8)&(9)&(10)&(11)&(12)\\
\hline  
           1   & $1.8$ &  0.2  &	 15.4  &   0.6    &  0.12    &  0.02  &   0.11 &  0.09   &  2.1  &    0.05	 &    9      \\ 
       2   & $4.8$ &  4.4  &	 13.9  &   0.7    &   0.11    &  0.03  &  0.07 &  0.09    &  1.9  &    0.06	  &  {\bf 21}  \\    
       3   & $1.6$ &  0.3  &	 12.6  &   0.5    &   0.11    &   0.02  &  0.07 &  0.05   &  --   &    --	  &	2	\\
       4   & $1.6$ &  0.5   &	 13.5  &   0.5    &   0.10    &  0.06  &  0.09 &  0.07    &  1.4  &    0.02	  &	8      \\  
       5   & $7.5$ &  4.5   &	 12.1  &   0.6    &   0.06    &  0.05  &  0.04 &  0.04    &  1.5  &    0.01	  &  {\bf 28  }  \\    
       6   & $16$  &  7.6   &	 11.8  &   0.9    &    0.12    &  0.04  &  0.02 &  0.01    &  1.7  &	0.04	   &  {\bf 152  }  \\	 
       7   & $5.6$ &  2.4   &	 12.1  &   0.5    &  0.09     &  0.04  &  0.03 &  0.02    &  1.6  &    0.03	  &  {\bf 22  }     \\    
       8   & $2.0$ &  0.7   &	 13.6  &   0.5    &   0.08    &   0.03  &   0.15 & 0.01   &  1.5  &    0.05	  &	6     \\  
       9   & $3.3$ &  1.3   &	 12.5  &   0.3    &  0.14     &  0.02  &  0.05 &  0.06    &  1.5  &    0.04	  &	10     \\  
      10   & $3.3$ &  0.6   &	  13.6  &   0.4    &   0.14    &  0.03  &  0.10 &  0.02    &  2.1  &	0.06	   &	  14	 \\  
      11   & $5.3$ &  2.5   &	 11.7  &   0.5    &  0.08     &  0.03 &  0.05 &  0.03	  &  1.9  &    0.04	 &     15   \\ 
      12   & $12$  &  4.1   &	 11.7  &   1.0    &    0.18    &  0.03  &  0.02 & 0.01     &  1.5  &	0.03	   &  {\bf  41  } \\	
      13   & $6.9$ &  5.8   &	 12.7  &   0.8    &   0.08    &  0.03 &  0.04 &  0.09	  &  1.6  &    0.02	  &  {\bf  23  }\\   
      \,\,\,\,\,$14^{\,\bf(d)}$    & $1.5$ &  0.8   &	 13.7  &   0.5    &    0.13   &  0.03  &   0.21 &  0.01   &  1.5--2.4  &    0.04--0.14	  &	 13 \\
      15   & $2.1$ &  2.2   &	 14.7  &    1.1    &  0.11    &   0.04  &   0.16 &  0.06  &  --   &	--	  &	 4 \\ 
          \,\,\,\,\,$16^{\,\bf(e)}$ 	 & $3.3$ &  0.9  &     26.9  &      2.1    &  0.12	 &  0.02  &   0.19 &  0.05   &  --   &     --	 &   12 \\ 
          \,\,\,\,\,$17^{\,\bf(e)}$ 	 & $2.0$ &  1.7  &     18.3  &      2.9    & 0.09	 &  0.05  &   0.20 &   0.29  &  1.3  &    0.04       &      6  \\   
          \,\,\,\,\,$18^{\,\bf(e)}$ 	 & $3.9$ &  2.0  &     22.2  &    2.6	 & 0.11 &  0.04  &   0.12 &  0.09   &  --   &	  --	&    {\bf  49	}   \\  
       \,\,\,\,\,$19^{\,\bf(e)}$    	 & $3.6$ &  1.5   &    26.5  &    1.2	 & 0.09   &  0.03  &   0.17 &  0.08   &  1.9  &    0.08   &	 1 \\	
      20   & $0.57$&  0.5    &    15.5  &    0.5   &  0.08     &  0.02  &   0.24 & 0.01    &  1.5  &	0.03	   &	  3 \\  
      21   & $1.5$ &  0.5   &	 14.0  &   0.4    &  0.13     &   0.03  &  0.09 &  0.06   &  1.7  &    0.05	  &	 6	\\   
       \,\,\,\,\,$22^{\,\bf(f)}$   & $0.8$ &  0.1   &	 13.5  &   0.3    &  $< 0.06$        &  0.19  &   0.32 &  0.05   &  --   &    --	  &	 1	\\    
      \,\,\,\,\,$23^{\,\bf(d)}$    & $1.4$ &  0.4   &	 13.2  &   0.7    &  0.14     &  0.07  &  0.09 &  0.07    &  1.5--1.8  &    0.04--0.06	  &	 7	 \\   
      24   & $5.6$ &  1.1   &	 12.3  &   0.2    &  0.07     &  0.03  &  0.05 & 0.01	  &  --   &	--	  &   {\bf 21	}    \\   
      25   & $2.2$ &  1.5  &	 12.7  &   0.5    &   0.10    &   0.03  &  0.09 & 0.01    &  1.5  &    0.03	  &	 11 \\
      \,\,\,\,\,$26^{\,\bf(d)}$    & $1.5$ &  0.3   &	 12.5  &   0.4    &  0.10     &  0.06  &   0.19 &  0.04   &  1.5--2.1  &   0.05-- 0.08	  &	 2	 \\  
      \,\,\,\,\,$27^{\,\bf(f)}$   & $2.9$ &  1.0   &	 11.5  &   0.3    &   $< 0.06$    &  0.03  &  0.03 &  0.03    &  1.4  &    0.02	  &    13  \\	\hline   
          mean   & 3   &1& 13.5& 0.5  &0.10&    0.03     &   0.09   &   0.05            & 1.6 &  0.03  & 11       \\     
         \hline         
         range  & 0.6-16&0.1--7.6 & 11--27 & 0.2--2.9 &  0.06--0.18 &  0.02--0.19&  0.02--0.32   &  0.01--0.29  & 1.3--2.4 & 0.01--0.08 & 1--152\\
         \hline  
         rms   & 3.4 &1.9 & 4.1&   0.7 &0.03  & 0.03 &    0.07  &      0.06         &0.3    &0.02& 30   \\
                           \hline  
                  \end{tabular}
\vspace*{-0.45ex}
 \begin{list}{}{}
 \item[]{{\bf Notes:} {Col.~2, Col.~4, Col.~8:} Mean projected column density, dust temperature, and Jeans length estimated at the center of each filament. 
{ Col.~6:} Weighted mean of the distribution of deconvolved FWHM widths resulting from Gaussian fits to the radial profiles at each position along the filament. 
 $^{\bf(a)}$ The quoted dispersion is the standard deviation of the {\it distribution} of the measured parameter along each filament. 
 In the case of $\langle\,{\rm FWHM}_{\rm dec}\rangle$ (Col.~6), this dispersion (Col.~7) is larger than the error bar plotted in Fig.~\ref{width_coldens} which corresponds 
 to the standard deviation of the {\it mean} deconvolved FWHM width. 
 $^{\bf(b)}$ $p$ and $R_{\rm flat}$ correspond to the best-fit model of the form expressed by Eq.~(1) to the mean column density profile of each filament. 
The fit was weighted by $1/\sigma(r)^2$ where $\sigma(r)$ is the standard deviation of the radial profile at projected radius $r$, estimated from the measured ($1\sigma$) dispersion along the filament. 
The same bounds (shown as vertical dashed lines in Figs.~\ref{fil_prof}a,~\ref{fil_prof}c,~\ref{fil_prof_14} for filaments 6 and 14) 
were used to fit the profiles and to calculate the mass per unit length by integrating over the radial profile.
 $^{\bf(c)}$ Col.~12 gives to the projected mass per unit length obtained by integrating over the observed column density profile of each filament. 
 The thermally supercritical filaments with $M_{\rm line} >  M_{\rm line, crit}$ are indicated in boldface, where $M_{\rm line, crit} = 2\, c_s^2/G \sim 20\, M_\odot$/pc  
(for a gas temperature $T \sim 12$~K, corresponding to the median central dust temperature measured here, ignoring the temperatures of the filaments located in the PDR region). 
  $^{\bf(d)}$ Filaments 14, 23, and 26 have asymmetric radial profiles. A separate fit was thus performed on either side of each of these three filaments. 
The values of $p$ and  $R_{\rm flat} $ reported in Col.~10 and Col.~11 correspond to the best-fit values on either side of these filaments.
   $^{\bf(e)}$ Filaments 16, 17, 18, and 19 are located in the PDR region called the {\it Cocoon Nebula} (see Fig.~\ref{coldens}a and online Fig.~\ref{color_image}), and
their derived dust temperatures may be overestimated. 
$^{\bf(f)}$ Filaments 22 and 27 have unresolved widths and are shown as upper limits corresponding to 75\% of the HPBW resolution in Fig.~\ref{width_coldens}.

}
 \end{list}      
\end{table*}
}


\onltab{2}{
\begin{table*}
\caption{Median values of observed angular widths and deconvolved physical widths$^{ \bf(a)}$ for three samples of filaments and two angular resolutions  }
\label{table}
\centering
\begin{tabular}{|c| c| cccc |ccc c| }
\hline
Field &  distance [pc] &  \multicolumn{4}{|c|}{From column density maps with 36.9\arcsec ~resolution} & \multicolumn{4}{|c|}{{\revised From SPIRE 250~$\mu$m images with 18.1\arcsec ~resolution} } \\
\hline
& & ${\rm FWHM}_{\rm obs}$ & $\sigma_{{\rm FWHM}_{\rm obs}}$  &  ${\rm FWHM}_{\rm dec}$ & $\sigma_{{\rm FWHM}_{\rm dec}}$ &    ${\rm FWHM}_{\rm obs}$ & $\sigma_{{\rm FWHM}_{\rm obs}}$  &  ${\rm FWHM}_{\rm dec}$ &$\sigma_{{\rm FWHM}_{\rm dec}}$ \\
& &  [\arcsec] &[\arcsec]&[pc] &  [pc] & [\arcsec] &[\arcsec]& [pc] & [pc]  \\
(1)&(2)&(3)&(4)&(5)&(6)&(7)&(8)&(9)&(10)\\
\hline
IC5146 &460 &59&15&0.10 & 0.03&44&10&0.09&0.02\\
Aquila &260 &94&21&0.11 & 0.03&97&31&0.12&0.04 \\
Polaris &150 &94&24&0.06 & 0.02&74&20&0.05&0.02\\
\hline
All$^{\bf(b)}$ & &  & &0.10 & 0.03& & &0.09&0.02\\
                \hline
         \end{tabular}
          \begin{list}{}{}
 \item[]{{\bf Notes:} {Col.~3 and Col.~7:} Median value of the observed FWHM angular width before deconvolution. 
 {Col.~5 and Col.~9:} Median value of the deconvolved FWHM physical width. 
 {Col.~4, Col.~6, Col.~8, Col.~10:} Dispersion (standard deviation) of the distribution of widths for each filament sample.  
 $^{\bf(a)}$ The DisPerSE algorithm \citep{Sousbie2010} we used to trace filaments in the column density maps derived from $Herschel$ data 
does not consider filamentary width in its process of identifying filaments. While it is difficult to assess the completeness of our census of filamentary structures 
without dedicated tests (which will be the subject of future work), there is in principle no bias toward selecting structures of similar width with DisPerSE.
$^{\bf(b)}$ The last row 
refers to the combined sample of  90 filaments observed in IC5146, Aquila, and Polaris.   
} 
 \end{list}        
\end{table*}
}

\onlfig{7}{
 \begin{figure*}
    \resizebox{0.5\hsize}{!}{\includegraphics[angle=0]{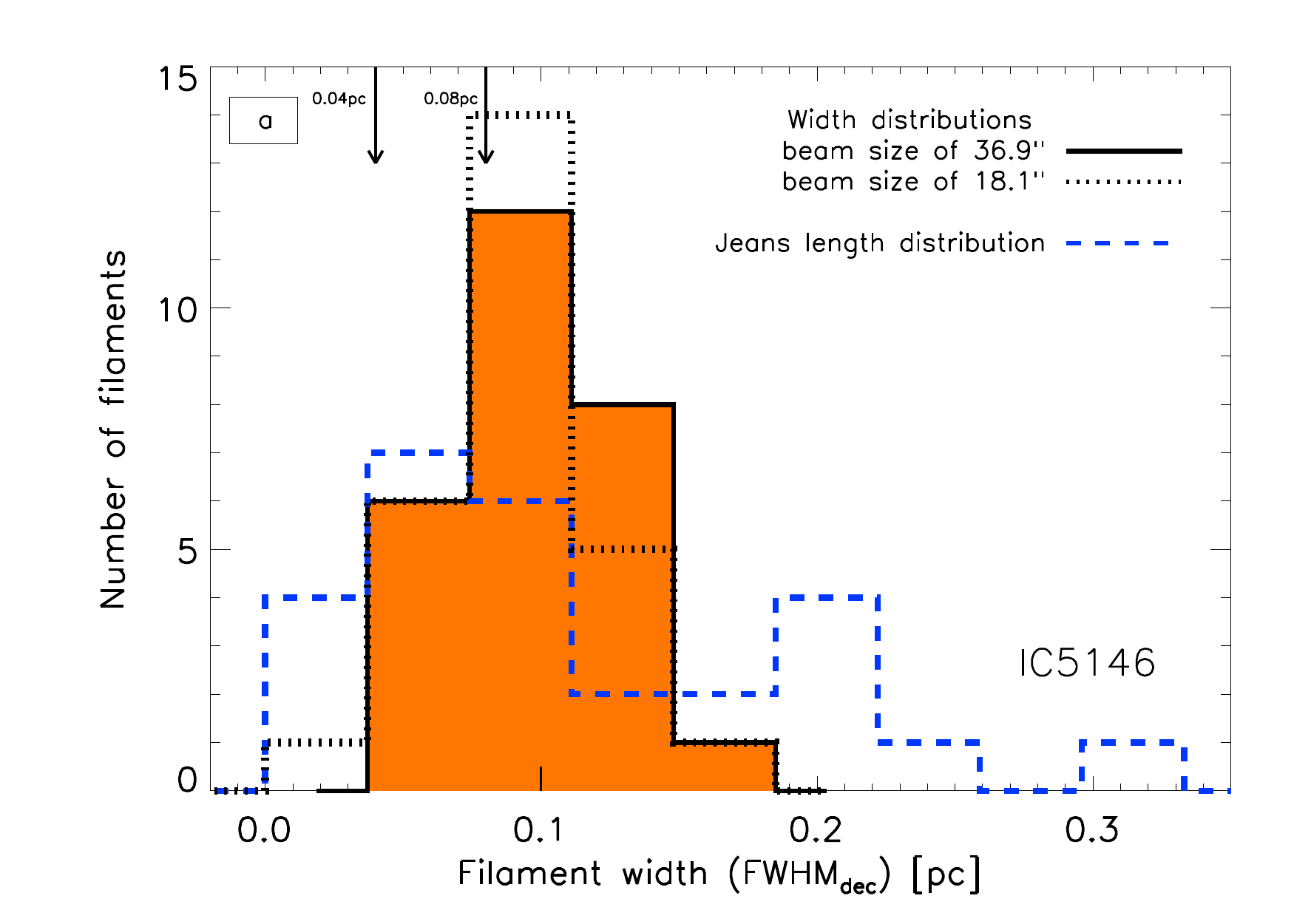}}
             \hspace{0.4mm}
     \resizebox{0.5\hsize}{!}{\includegraphics[angle=0]{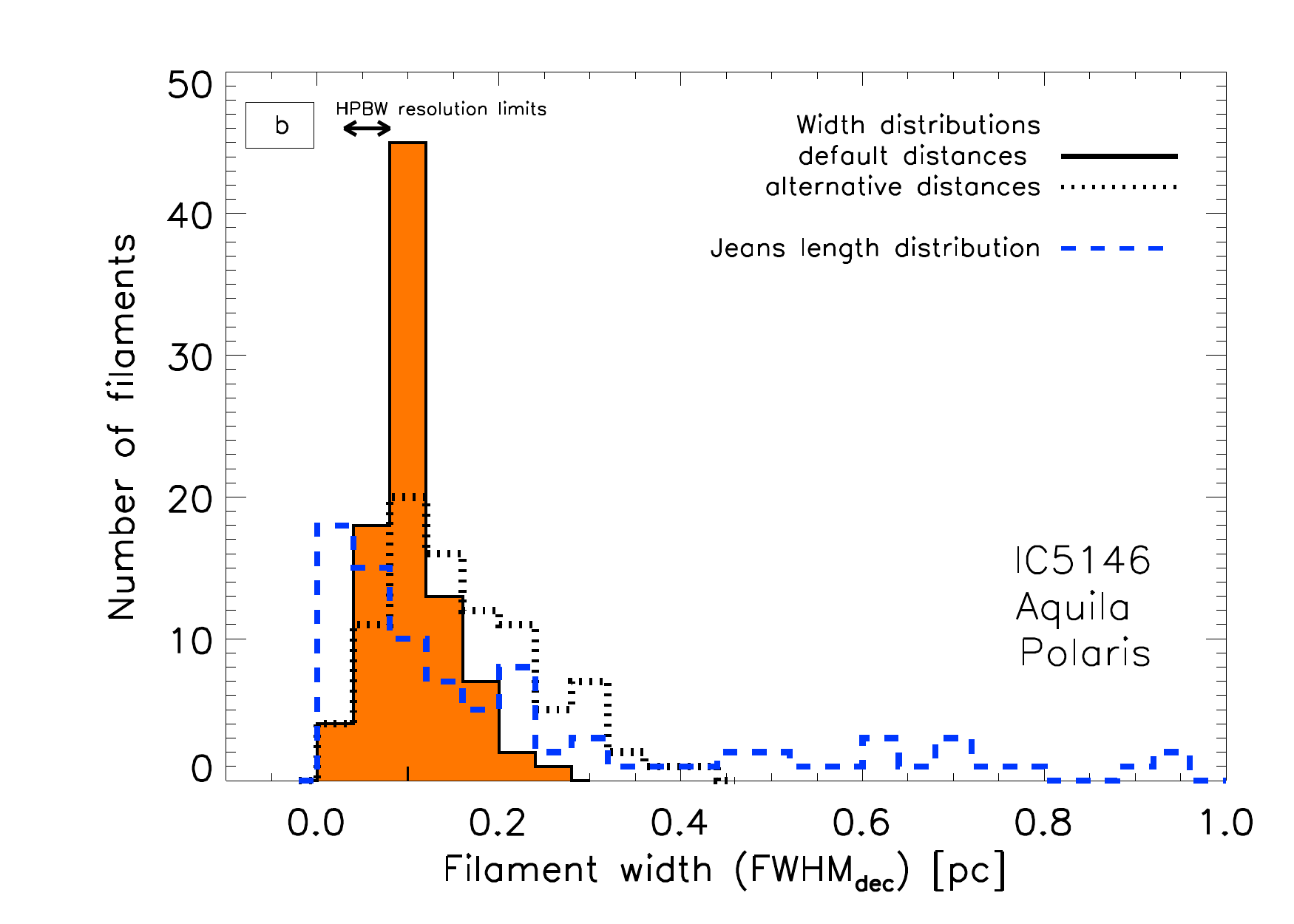}}
   \caption{{\bf (a)}  Distribution of deconvolved FWHM widths for the 27 filaments of IC5146 (black solid histogram, filled in orange).
These widths have been deconvolved from the $36.9\arcsec $~HPBW resolution of the column density map (Fig.~\ref{coldens}a)
used to construct the radial profiles of the filaments.
The median width is 0.10~pc and the standard deviation of the distribution is $0.03~$pc.
The black dotted histogram shows the distribution of deconvolved FWHM widths measured on 
{\revised 
the SPIRE 250~$\mu$m map which has a factor $\sim 2$ better resolution (18.1\arcsec ~HPBW).
}. 
This distribution is statistically indistinguishable from the distribution obtained at $36.9\arcsec $ resolution. 
{\revised 
The two down pointing arrows in the upper left mark the resolution limits for the distributions at $36.9\arcsec \,$ and 18.1\arcsec ,  i.e., 0.08~pc and 0.04~pc, respectively. 
}
For comparison, the blue dashed histogram represents the distribution of central Jeans lengths corresponding
to the central column densities of the filaments.
{\bf (b)} Distributions of deconvolved FWHM widths for a larger sample of 90 filaments, combining the 27 filaments of IC5146,
32 filaments in Aquila, and 31 filaments in Polaris, all analyzed in the same way from column density maps with $36.9\arcsec $ resolution 
as explained in Sect.~3 and Sect.~4.
The black solid histogram, filled in orange, is based on our default distance assumptions: 460~pc for IC5146, 260~pc for Aquila, and 150~pc for Polaris.  
This distribution has a median of 0.10~pc and a standard deviation of $0.03~$pc.
{\revised 
The horizontal double arrow at the upper left 
shows the range of HPBW resolution limits, going from $\sim 0.03$~pc for Polaris to $\sim 0.08$~pc for IC5146. 
}  
The dotted histogram shows an alternate distribution of deconvolded widths for the same sample of filaments based on other distance assumptions: 950~pc for IC5146 (see Appendix~A), 
400~pc for Aquila \citep[see discussion in][and Appendix of Andr\'e et al. 2010]{Bontemps2010}, and 150~pc for Polaris. 
The median value of this alternate distribution is 0.15~pc and the standard deviation is 0.08~pc. 
For comparison, the blue dashed histogram represents the distribution of central Jeans lengths corresponding to the central column densities of the 90 filaments 
(independent of distance).
 }
              \label{histo}%
    \end{figure*}
    }

\bibliography{AA}

\Online
\begin{appendix} 
\section{Effects of distance and dust opacity uncertainties and influence of the viewing angle}

There is some ambiguity concerning the distance to IC5146. The default distance adopted in this paper corresponds 
to the value of $460^{+40}_{-60}Ê$~pc derived from star counts by \citet{Lada1999}. 
However, other studies placed the IC5146 cloud at a distance of $950 \pm 80 Ê$~pc based on photometry and spectra 
of late-B stars in the IC5146 cluster \citep[e.g.][]{Harvey2008}. 
If the true distance of IC5146 were $\sim 950$~pc instead of $\sim 460$~pc, then the widths of the IC5146 filaments 
would all be a factor of $\sim 2$ larger than the values listed in Col.~6 of Table~1, leading to a median width of z
$0.2 \pm 0.06$~pc instead of $0.1 \pm 0.03$~pc. 
This would also broaden the distribution of FWHM widths for the combined sample of 90 filaments in IC5146, Aquila, 
and Polaris (see dotted histogram in online Fig.~\ref{histo}b). 
As can be seen in online Fig.~\ref{histo}b, our main result that the typical filament width is $\sim 0.1$~pc to within 
a factor of $\sim 2$ would nevertheless remain valid. 
We also stress that the distance uncertainty has no effect on the {\it shape} of the radial column density 
profiles (Sect.~3.2 and Fig.~\ref{fil_prof}) or on the absence of an anti-correlation between filament width 
and central column density in Fig.~\ref{width_coldens}. 

{\revised 
The following (temperature-independent) dust opacity law was assumed in our analysis of the $Herschel $ data:\\  
$\kappa_{\nu} = 0.1 \times \left(\nu/1000~{\rm GHz}\right)^2$~cm$^2$/g, where $\nu $ denotes frequency and 
$\kappa_{\nu}$ is the dust opacity per unit (gas~$+$~dust) mass column density.
This dust opacity law is very similar to that advocated by \citet{Hildebrand1983} at submillimeter wavelengths, and 
is consistent with the mean value $\kappa_{\rm 850\mu } \approx 0.01$~cm$^2$/g derived by \citet{Kramer2003} 
in IC5146 from a detailed comparison of their SCUBA 850~$\mu$m and 450~$\mu$m data with the near-infrared extinction 
map of \citet{Lada1999}. However, \citet{Kramer2003} found evidence of an increase in the dust opacity $\kappa_{\rm 850\mu} $ 
by a factor of $\sim 4 $ when the dust temperature $T_d$ decreased from $\sim 20$~K to $\sim 12$~K, 
which they interpreted as the manifestation of dust grain evolution (e.g. coagulation and formation of ice mantles) in the cold, dense interior of the cloud. 
The dust temperature map derived from $Herschel$ data (see Fig.~\ref{250_curv_crit}a) suggests that 
$T_d$ ranges from $\sim 11$~K to $\sim 17$~K in the bulk of IC5146, with the exception of the PDR region associated 
with the Cocoon Nebula where $T_d$ reaches $\sim 30$~K. 
Assuming a linear increase in $\kappa_{\nu}$ by a factor of 4 when $T_d$ decreases from $\sim 20$~K to $\sim 12$~K
as suggested by the \citet{Kramer2003} study, we estimate that the column density map shown in 
Fig.~\ref{coldens}a is accurate to better than a factor of $\sim 2$ over most of its extent. 
The possible dependence of $\kappa_{\nu}$ on temperature has very little impact on our analysis 
of the filament profiles. For filament 6, for instance, $T_d$ decreases by less than $\sim 3$~K 
from the exterior to the interior of the filament (see Fig.~\ref{fil_prof}b), suggesting that 
$\kappa_{\nu}$ does not change by more than $\sim 40\% $ from large to small radii. 
The potential effect on the estimated FWHM width, $W$, and power-law index, $p$, of the filament profiles  
is even smaller: $W$ would increase, and $p$ would decrease, by less than $\sim 3\%$ and $\sim 7\%$, respectively.

}

Our $Herschel$ observations only provide information on the {\it projected} column density profile   
$\Sigma_{\rm obs}(r) = \frac{1}{{\rm cos} \,i} \, \Sigma_{\rm int}(r)$ of any given filament, where $i$ is the inclination angle to the plane of the sky 
and $\Sigma_{\rm int}$ is the intrinsic column density profile of the filament. 
For a population of randomly oriented filaments with respect to the plane of the sky, 
the net effect is that $\Sigma_{\rm obs} $ {\it overestimates} $\Sigma_{\rm int} $ by a factor  
$<\frac{1}{{\rm cos}\,i}>\, = \frac{\pi}{2}\,\sim1.57$ on average. 
This does not affect our analysis of the shape of the radial column density profiles (Sect.~3.2), but implies that the central column densities 
of the filaments are actually $\sim 60\% $ lower on average than the projected values listed in Col.~2 of Table~1. 
Likewise, the observed masses per unit length (Col.~12 of Table 1) tend to overestimate the true masses per unit length of the filaments 
by $\sim 60\% $ on average. 
Although systematic, this inclination effect remains less than a factor of 2 and thus has little impact on the classification of 
observed filaments in thermally subcritical and thermally supercritical filaments.

\label{app_cosi}
\end{appendix}


%
%

 
\end{document}